\documentclass[aoas,nameyear,dvips]{arximspdf}
\usepackage{dcolumn}
\usepackage{graphicx}

\doi{10.1214/10-AOAS376}
\volume{5}
\issue{1}
\pubyear{2011}
\firstpage{551}
\lastpage{577}

\makeatletter
\newcolumntype{d}[1]{D{.}{.}{#1}}
\makeatother

\begin{document}
\begin{frontmatter}

\title{A dynamic Bayesian nonlinear mixed-effects model of HIV
response incorporating medication adherence, drug resistance and
covariates\thanksref{T1}}

\runtitle{Dynamic Bayesian nonlinear mixed-effects model}

\thankstext{T1}{Supported in part by NIAID/NIH Grant AI080338 and MSP/NSA Grant
H98230-09-1-0053 to Y. Huang, and NIH Grants AI50020, AI078498, AI078842 and AI087135 to H. Wu.}

\begin{aug}
\author[A]{\fnms{Yangxin} \snm{Huang}\corref{}\ead[label=e1]{yhuang@health.usf.edu}},
\author[B]{\fnms{Hulin} \snm{Wu}\ead[label=e2]{hwu@bst.rochester.edu}},
\author[B]{\fnms{Jeanne} \snm{Holden-Wiltse}\ead[label=e3]{Jeanne\_Wiltse@URMC.Rochester.edu}}
\and\\
\author[C]{\fnms{Edward P.} \snm{Acosta}\ead[label=e4]{Edward.Acosta@ccc.uab.edu}}

\runauthor{Huang et al.}

\affiliation{University of South Florida, University of Rochester,
University of Rochester and University of Alabama}

\address[A]{Y. Huang\\
Department of Epidemiology\\
and Biostatistics\\
College of Public Health, MDC 56\\
University of South Florida\\
Tampa, Florida 33612\\
USA\\
\printead{e1}}

\address[B]{H. Wu\\
J. Holden-Wiltse\\
Department of Biostatistics\\
and Computational Biology\\
School of Medicine and Dentistry\\
University of Rochester\\
Rochester, New York 14642\\
USA\\
\printead{e2}\\
\phantom{E-mail: }\printead*{e3}}

\address[C]{E. P. Acosta\\
Birmingham School of Medicine\\
University of Alabama \\
Division of Clinical Pharmacology\\
1530 3rd Avenue South\\
Birmingham, Alabama 35294\\
USA\\
\printead{e4}}
\end{aug}

\received{\smonth{5} \syear{2009}}
\revised{\smonth{9} \syear{2009}}


\begin{abstract}
\fontsize{9}{10.5}\selectfont{
HIV dynamic studies have contributed significantly
to the understanding of HIV pathogenesis and antiviral treatment
strategies for AIDS patients. Establishing the relationship of
virologic responses with clinical factors and covariates during
long-term antiretroviral (ARV) therapy is important to the
development of effective treatments. Medication adherence is an
important predictor of the effectiveness of ARV treatment, but an
appropriate determinant of adherence rate based on medication event
monitoring system (MEMS) data is critical to predict virologic
outcomes. The primary objective of this paper is to investigate the
effects of a number of summary determinants of MEMS adherence rates
on virologic response measured repeatedly over time in HIV-infected
patients. We developed a mechanism-based differential equation
model with consideration of drug adherence, interacted by virus
susceptibility to drug and baseline characteristics, to characterize
the long-term virologic responses after initiation of therapy. This
model fully integrates viral load, MEMS adherence, drug resistance
and baseline covariates into the data analysis. In this study we
employed the proposed model and associated Bayesian nonlinear
mixed-effects modeling approach to assess how to efficiently use the
MEMS adherence data for prediction of virologic response, and to
evaluate the predicting power of each summary metric of the MEMS
adherence rates. In particular, we intend to address the questions:
(i)~how to summarize the MEMS adherence data for efficient
prediction of virologic response after accounting for potential
confounding factors such as drug resistance and covariates, and (ii)
how to evaluate treatment effect of baseline characteristics
interacted with adherence and other clinical factors. The approach
is applied to an AIDS clinical trial involving 31 patients who had
available data as required for the proposed model. Results
demonstrate that the appropriate determinants of MEMS adherence
rates are important in order to more efficiently predict virologic
response, and investigations of adherence to ARV treatment would
benefit from measuring not only adherence rate but also its summary
metric assessment. Our study also shows that the mechanism-based
dynamic model is powerful and effective to establish a relationship
of virologic responses with medication adherence, virus resistance
to drug and baseline covariates.}
\end{abstract}

%

\end{frontmatter}
%

\section{Introduction} \label{sec1}
The revolution in human immunodeficiency virus (HIV)
treatment has brought diagnostic tests that can accurately measure
levels of HIV in blood. Resulting data show (plasma) viral load
(HIV-1 RNA copies or RNA copies) to be an important predictor of the
risk of progression to AIDS. The antiretroviral (ARV) agents, which
include potent protease inhibitors (PIs) are, however, not a cure
for HIV infection. While many patients benefit from ARV treatment,
others do not benefit or only experience a temporary benefit. There
are several reasons why treatment fails, of which poor patient
adherence to ARV therapy is a leading factor [Ickovics and Meisler
(\citeyear{Ickovics97bib19}); Paterson {et al.} (\citeyear{Paterson00bib26-1})]. Thus, assessment of adherence within
AIDS clinical trials is a critical component of the successful
evaluation of therapy outcomes. Maintaining adherence may be
particularly difficult when the drug regimen is complex or
side-effects are severe, as is often the case for current HIV
therapy [Ickovics and Meisler (\citeyear{Ickovics97bib19})].

The measurement of adherence remains problematic; a standard
\mbox{definition} of adherence and completely reliable measures of
adherence are lacking. Never\-theless, there has been substantial
progress in both of these areas in the past few years. First, it
appears that higher levels of adherence are needed for HIV disease
than other diseases to achieve the desired therapeutic benefit
[Paterson et al. (\citeyear{Paterson00bib26-1})]. Second, better appreciation of the value
and limitations of different adherence measurements has been
addressed [Bova et al. (\citeyear{Bova05bib5-2})]. In AIDS clinical trials adherence to
medication regimen is currently measured by two methods: by use of
questionnaires (patient self-reporting and/or face-to-face
interview) and by use of electronic compliance monitoring
(Medication Event Monitoring System [MEMS]) caps. The MEMS is often
used as an objective adherence measure. It consists of a computer
chip in the cap of a medication bottle that records each time the
bottle is opened. The results can be downloaded, printed out and
analyzed. It demonstrates that medication-taking patterns are highly
variable among patients [Kastrissios et al. (\citeyear{Kastrissios98bib20-2})] and that they
often give a more precise measure of adherence than self-report
[Arnsten et al. (\citeyear{Arnsten01bib3})]. However, MEMS data are~also subject to error
and are not widely available in the clinical setting. Adherence
assessment by self-report is usually evaluated by a patient's
ability to recall their medication dosing during a specific time
interval. Finally, it is important to note that the measurement of
viral load levels is of special utility as an indirect measure of
adherence in HIV therapeutics. It has been argued that this is not
an adherence measure because other factors may influence viral load
(drug resistance, etc.). However, there is a tight correlation
between
viral load and adherence [Paterson et al. (\citeyear{Paterson00bib26-1}); Haubrich et al.
(\citeyear{Haubrich99bib20-4})].

Viral dynamic models can be formulated through ordinary differential
equations (ODE), but there has been only limited development of
statistical methodologies for assessing their agreement with
observed data. Currently there also are substantial knowledge gaps
between theoretical HIV dynamics and the role of many clinical
factors. In developing long-term dynamic modeling, this paper will
address these problems by utilizing time-specific information, such
as drug adherence and susceptibility factors, on the biological
mechanism of HIV dynamics to achieve more realistic and accurate
characterization of the relationship between clinical/drug factors
and virologic response. Several studies [Arnsten et al. (\citeyear{Arnsten01bib3});
Levine et al. (\citeyear{Levine06bib21-1})] investigated the association between virologic
responses and adherence assessed by both questionnaire and MEMS
data. The results indicated that the MEMS cap adherence data may not
be correlated better to virologic response compared to the
questionnaire adherence data unless the MEMS cap data are summarized
in an appropriate way. Further, Huang et al. (\citeyear{Huang08bib18-2}), Labb\'{e}
and Verotta (\citeyear{Labbe06bib21}), Liu et al. (\citeyear{Liu07bib21-2})
and Vrijens et al. (\citeyear{Vrijens05bib38})
modeled the relationship between virologic response and adherence
rate using questionnaire data and MEMS data averaged by each
interval between study visits or weekly basis, but no significant
differences were found in predicting virologic response. Along with
this line, this paper will investigate different determinants of
the adherence rate based on MEMS data from an AIDS clinical trial study
[Hammer et al. (\citeyear{Hammer02bib13})] and compare their performance for predicting
a virological response. We employed the proposed mechanism-based
dynamic model to assess how to efficiently use the adherence data
based on MEMS to predict virological response. In particular, we
intend to address the questions (i) how to summarize the MEMS
adherence data for efficient prediction of virological response
after accounting for potential confounding factors such as drug
resistance and baseline covariates, and (ii) how to evaluate
treatment effect of baseline characteristics interacted with MEMS
adherence and other clinical
factors.

The purpose of this paper is to describe a reparameterized ODE
dynamic model (with identifiable parameters) which fully integrates
viral load, medication adherence, drug
resistance and baseline covariates data from an AIDS clinical trial
study into the analysis. Thus, our dynamic model will be able to
characterize sustained suppression or resurgence of the virus as
arising from intrinsic viral dynamics, and/or influenced by factors
such as drug susceptibility and adherence during the treatment
period of the clinical trial. The Bayesian nonlinear mixed-effects
(BNLME) modeling approach [Davidian and Giltinan (\citeyear{Davidian95bib8})] is employed
to estimate dynamic parameters and identify significant clinical
factors and/or covariates on virologic response to ARV treatment.
The rest of this article is organized as follows. Section \ref{sec2}
introduces reparameterized viral dynamic models with time-varying
drug efficacy which incorporates the effects of drug adherence, drug
resistance and baseline covariates, and briefly describes the BNLME
modeling approach, implemented via Markov chain Monte Carlo (MCMC)
procedures, followed by defining the deviance information criterion
(DIC) for comparison of models. In Section \ref{sec3} we summarize
the motivating data set from an AIDS clinical trial study including
the data of plasma viral load, medication adherence from MEMS cap,
drug resistance and baseline covariates; the proposed methodology is
applied to these data and the results are presented. The method is
evaluated via a simulation study in Section \ref{sec4}. Finally, we
conclude the article with some discussions in Section \ref{sec5}.

\section{HIV dynamic mechanism-based ODE models and statistical
approaches} \label{sec2}
This section aims to introduce long-term viral dynamic models
based on a system of ODE with time-varying coefficients but without
closed-form solutions, and to investigate associated methodologies
to demonstrate the application of these models to an AIDS clinical
trial study. Long-term viral dynamic models can be used to describe
the interaction between cells susceptible to target cells ($T$),
infected cells ($T^*$) and free virus ($V$) by considering
time-varying drug efficacy [Huang and Wu (\citeyear{Huang06bib17}); Huang et al. (\citeyear{Huang06bib18})].
These three compartments (variables) are described as follows.

HIV virions ($V$) will infect target cells ($T$) and
turn them into infected cells ($T^*$) at an infection rate $k$. Due
to the intervention of antiviral drugs, we assume that drugs reduce
the infection rate in the infected cells ($T^*$) by $1-\gamma(t)$
[$0\leq\gamma(t)\leq1$]. The infected cells will die at rate
$\delta$ after producing an average of $N$ virions per cell during
their lifetimes, and free virions are removed from the system at
rate $c$. In addition to the dynamics describing virus infection, we
have to specify the dynamics of the uninfected cell population. The
simplest assumption is that uninfected cells are produced at a
constant rate $\lambda$ at which new $T$ cells are generated from
sources within the body, such as the thymus and die at a rate $d_T$.
Thus, the HIV dynamic model, after initiation of antiviral therapy,
can be written as
\begin{eqnarray}\label{ode1}
\frac{d}{dt} T(t)&=& \lambda-d_T T(t)-[1 -\gamma(t)]k T(t) V(t),\nonumber\\
\frac{d}{dt} T^*(t) &=& [1 - \gamma(t)]k T(t) V(t) - \delta
T^*(t) ,\\
\frac{d}{dt} V(t) &=& N \delta T^*(t)-c V(t),\nonumber
\end{eqnarray}
where the time-varying parameter $\gamma(t)$ (as defined
below) quantifies the time-varying drug efficacy. If the regimen is
not 100\% effective [i.e., $0\leq\gamma(t)<1$], the system of ODE
cannot be solved analytically. The solutions to (\ref{ode1}) then
have to be evaluated numerically. When $\gamma(t)=\gamma_0$ (an
unknown constant), the model~(\ref{ode1}) becomes the model
developed by Perelson and Nelson (\citeyear{Perelson99bib29}). In particular, when
$\gamma(t)=0$ (the drug has no effect), the model (\ref{ode1})
reduces to the model in the publications [Bonhoeffer et al. (\citeyear{Bonhoeffer97Abib5-1});
Nowak et al. (\citeyear{Nowak95bib24,Nowak97bib24-1,Nowak00bib25}); Stafford et al. (\citeyear{Stafford00bib34})]; while
$\gamma(t)=1$ (the drug is 100\% effective), the model (\ref{ode1})
reverts to the model discussed by
Nowak and May (\citeyear{Nowak00bib25}) and Perelson and Nelson (\citeyear{Perelson99bib29}).

However, it is challenging to estimate all the parameters in
the model (\ref{ode1}) and to conduct inference because the model
(\ref{ode1}) is not a priori identifiable (i.e., multiple sets of
parameters obtain identical fits to the data), given only viral load
measurements [Cobelli et al. (\citeyear{Cobelli79bib7})]. To obtain a model with a priori
identifiable parameters [Labb\'{e} and Verttoa (\citeyear{Labbe06bib21})], this
paper investigates mechanism-based reparameterized ODE models to
quantify the long-term viral dynamics with ARV treatment and the
associated statistical methods for model fitting.

\subsection{Reparameterized model with time-varying drug
efficacy} \label{sec2.1}
Following the studies [Perelson and Nelson (\citeyear{Perelson99bib29}); Nowak and May (\citeyear{Nowak00bib25});
Labb\'{e} and Verttoa (\citeyear{Labbe06bib21})], we reparameterize the model
(\ref{ode1}) using the rescaled variables
$\widetilde{T}(t)=(d_T/\lambda)T(t)$,
$\widetilde{T}^*(t)=(\delta/\lambda)T^*(t)$,
$\widetilde{V}(t)=(k/d_T)V(t)$. These yield the rescaled version as
follows:
\begin{eqnarray}\label{ode2}
{d \over dt} \widetilde{T}(t)&=& \frac{d_T}{\lambda}{d \over
dt}T=d_T\{
1-\widetilde{T}(t)-[1 -\gamma(t)] \widetilde{T}(t) \widetilde
{V}(t)\},\nonumber\\
{d \over dt} \widetilde{T}^*(t) &=& \frac{\delta}{\lambda}{d \over dt}
T^*(t)= \delta\{[1 - \gamma(t)]\widetilde{T}(t) \widetilde{V}(t) -
\widetilde{T}^*(t)\}, \\
{d \over dt} \widetilde{V}(t) &=& \frac{k}{d_T}{d \over dt}V(t)=c\{R
\widetilde{T}^*(t)- \widetilde{V}(t)\},\nonumber
\end{eqnarray}
where $R=kN\lambda/(c d_T)$ represents the basic reproductive
ratio for the virus, defined as the number of newly infected cells
that arise from any one infected cell when almost all cells are
uninfected [Nowak and May (\citeyear{Nowak00bib25})]. Note that the rescaled model
(\ref{ode2}) has fewer parameters than the `original' model
(\ref{ode1}). The identifiability of the model (\ref{ode2}) is
guaranteed [Cobelli et al. (\citeyear{Cobelli79bib7}); Labb\'{e} and Verttoa (\citeyear{Labbe06bib21})]
and parameters of the model can be uniquely identified. If $R<1$,
then the virus will not spread, since every infected cell will on
average produce less than one infected cell. If, on the other hand,
$R>1$, then every infected cell will on average produce more than
one newly infected cell and the virus will proliferate. For the HIV
virus to persist in the host, infected cells must produce at least
one secondary infection, and $R$ must
be greater than unity [Nowak and May (\citeyear{Nowak00bib25})].

Assuming steady state before the beginning of drug therapy, initial
conditions for the model can now be expressed as simple functions of
the initial conditions for viral load ($\widetilde{V}_0$):
$\widetilde{T}_0=1/(1+\widetilde{V}_0),
\widetilde{T}_0^*=\widetilde{V}_0/(1+\widetilde{V}_0),
\widetilde{V}_0=\widetilde{V}(0)$ [Cobelli {et al.} (\citeyear{Cobelli79bib7});
Labb\'{e} and Verttoa (\citeyear{Labbe06bib21})]. The assumption of initial steady
state is necessary to guarantee identifiable (none of the models
reported or referenced here is identifiable if the initial states
are unknown), and is often justified by the clinical trial protocol.
For example, in ACTG 398, individual patients were taken off the drug
before the initiation of the new therapy (washout period to
eliminate the effect of previously administered drugs and to
guarantee that all individuals started from steady-state
conditions). Finally, viral load [$V(t)$] in model (\ref{ode1}) is
related to an equation output of viral load amount
[$\widetilde{V}(t)$] in model (\ref{ode2}) as follows: $ V(t)= \rho
\widetilde{V}(t)$, where $\rho$, which is equivalent to a volume of
distribution of pharmacokinetics, is a viral load scaling
(proportionality) factor (10,000 copies/ml) to be estimated from the
data [Nowak and May (\citeyear{Nowak00bib25})]. The set of ODE (\ref{ode2}) will be used
to construct the BNLME model.

\subsection{Time-varying drug efficacy model} \label{sec2.2}
Within the population of HIV virions in a human host, there is
likely to be genetic diversity and corresponding diversity in
susceptibility to the various ARV agents. In clinical practice,
genotypic or phenotypic tests can be performed to determine the
sensitivity of HIV-1 RNA to ARV agents before a treatment regimen is
selected. Here we use the phenotypic marker, $\mathit{IC}_{50}$ [Molla et
al. (\citeyear{Molla96bib23})], to quantify agent-specific drug susceptibility. Because
experimental data tracking development of resistance suggest that
the resistant fraction of the viral population grows exponentially,
we propose a model of $\log$-linear function to approximate
the within-host changes over time in $\mathit{IC}_{50}$ as follows:
\[
\mathit{IC}_{50}(t)=\cases{
\log\biggl(S_{0}+\displaystyle\frac{S_{f}-S_{0}}{t_{f}}t\biggr) & \quad $\mbox{for
$0<t<t_{f}$}$,\vspace*{2pt}\cr
\log(S_{f})& \quad $\mbox{for $t\geq t_{f}$},$}
\]
where $S_{0}$ and $S_{f}$ are respective values of $\mathit{IC}_{50}(t)$
at baseline and time point $t_{f}$ at which the resistant mutations
dominate. In our study, $t_{f}$ is the time of virologic failure
which is observed from clinical studies. If $S_{f}=S_{0}$, no new
drug resistant mutation is developed during treatment. Although more
complicated models for median inhibitory concentration have been
proposed based on the frequencies of resistant mutations and
cross-resistance patterns [Wainberg et al. (\citeyear{Wainberg96bib40}); Bonhoeffer, Lipsitch and Levin
(\citeyear{Bonhoeffer97bib5})], in clinical studies or clinical practice it is common to
collect $\mathit{IC}_{50}$ values only at baseline and failure time as
designed in ACTG 5055 [Acosta et al. (\citeyear{Acosta04bib1})] and ACTG 398 [Hammer et
al. (\citeyear{Hammer02bib13}); Pfister et al. (\citeyear{Pfister03bib30})]. Thus, given that $\mathit{IC}_{50}$ is only
measured at baseline and at the time of treatment failure, this
function may serve as a good approximation in terms of data
availability.

Poor adherence to a treatment regimen is one of
the major causes of treatment failure [Ickovics and Meisler (\citeyear{Ickovics97bib19})].
The following model is used to represent adherence
for a time interval $\textsl{T}_{k}<t \leq\textsl{T}_{k+1}$:
\[
A(t)=\cases{
1, & \quad $\mbox{if all doses are taken in ($\textsl{T}_{k}, \textsl
{T}_{k+1}$]},$\vspace*{2pt}\cr
r_k, &\quad $ \mbox{if $100 r_k\%$ doses are taken in ($\textsl{T}_{k},
\textsl
{T}_{k+1}$]},$}
\]
where $0\leq r_k<1$, with $r_k$ indicating the adherence rate
computed for each assessment interval ($T_k, T_{k+1}$] between study
visits based on the questionnaire or MEMS data;
$\textsl{T}_{k}$ denotes the $k$th adherence assessment time.

In most viral dynamic studies, investigators assumed that either
drug efficacy was constant over treatment time [Perelson and Nelson
(\citeyear{Perelson99bib29}); Wu and Ding (\citeyear{Wu99bib45})] or antiviral regimens had perfect effect in
blocking viral replication [Ho et al. (\citeyear{Ho95bib15}); Perelson et al. (\citeyear{Perelson96bib27})].
However, the drug efficacy may change as concentrations of ARV drugs
and other factors (e.g., drug resistance) vary during
treatment. A simple pharmacodynamic sigmoidal $E_{\mathrm{max}}$ model for
dose--effect relationship is [Gabrielsson and Weiner
(\citeyear{Gabrielsson00bib9})]
\begin{equation}
E= \frac{E_{\mathrm{max}}C}{EC_{50}+C},
\end{equation}
where $E_{\mathrm{max}}$ is the maximal effect that can be achieved, $C$
is the drug concentration, and $EC_{50}$ is the drug concentration that
induced an effect equivalent to $50\%$ of the maximal effect. Many
different variations of the $E_{\mathrm{max}}$ model have been developed by
pharmacologists to model pharmacodynamic effects. $E_{\mathrm{max}}$ models
include the sigmoid $E_{\mathrm{max}}$ model, the ordinary $E_{\mathrm{max}}$ model
and composite $E_{\mathrm{max}}$ models [Gabrielsson and Weiner (\citeyear{Gabrielsson00bib9});
Davidian and Giltinan (\citeyear{Davidian95bib8})]. The ordinary $E_{\mathrm{max}}$ model describes
agonistic and antagonistic (inhibitory) effects of a drug, the
sigmoid $E_{\mathrm{max}}$ model is more flexible for the steepness or
curvature of the response--concentration curve compared to the
ordinary $E_{\mathrm{max}}$ model, and composite $E_{\mathrm{max}}$ models are used
for multiple drug effects. More detailed discussions on $E_{\mathrm{max}}$
models can be found in the book by Gabrielsson and Weiner (\citeyear{Gabrielsson00bib9}) and
the paper by Huang et al. (\citeyear{Huang03bib16}). Here we employ the following
modified $E_{\mathrm{max}}$ model to represent the time-varying drug efficacy
for two ARV agents within a class,
\begin{equation}
\label{eff}
\gamma(t) =\frac{A_1(t)/\mathit{IC}_{50}^1(t)+ A_2(t)/\mathit{IC}_{50}^2(t)}
{\phi+A_1(t)/\mathit{IC}_{50}^1(t)+ A_2(t)/\mathit{IC}_{50}^2(t)},
\end{equation}
where $A_k(t)$ and $\mathit{IC}_{50}^k(t)$ $(k=1,2)$ are the adherence
profile of the drug as measured by MEMS data and the time-course of
median inhibitory concentrations for the two drugs, respectively;
$\phi=\exp(\beta_0+\beta_1 w_1+\beta_2 w_2)$; $w_1$ and $w_2$ are
observed baseline viral load and CD4 cell count, respectively;
$\bolds{\beta}=(\beta_0, \beta_1, \beta_2)^T$ are unknown
covariate effect
parameters to be estimated from clinical data. If
$\beta_1=\beta_2=0$ (without considering effect of covariates),
$\phi=\exp(\beta_0)$ can be used to quantify the conversion between
in vitro and in vivo $\mathit{IC}_{50}$ which is the case discussed by Huang
{et al.} (\citeyear{Huang03bib16}). If $\gamma(t)=1$, the drug is $100\%$ effective,
whereas if $\gamma(t)=0$, the drug has no effect. Note that, if
$A_k(t)$, $\mathit{IC}_{50}^k(t)$, $w_1$ and $w_2$ are measured or obtained
from a clinical study and $\bolds{\beta}$ can be estimated from clinical
data, then the time-varying drug efficacy $\gamma(t)$ can be
estimated for the whole period of ARV treatment.

\subsection{Bayesian modeling approaches} \label{sec2.3}
A number of studies investigated various statistical
methods, including Bayesian approaches, to fit viral dynamic models and
to predict virological responses [Han et al. (\citeyear{Han02bib14}); Huang et al. (\citeyear{Huang06bib18});
Perelson et al. (\citeyear{Perelson96bib27}); Wu et al. (\citeyear{Wu98bib44}); Wu and Ding (\citeyear{Wu99bib45})].
The Bayesian approach to viral dynamic modeling is particularly
appealing from a biological perspective, as it allows informative
prior distributions to be incorporated. From a statistical estimation
point of view, a Bayesian approach is preferable because of the
difficulties which are often encountered from a classical approach
when models involve the large numbers of parameters, and complex
nonlinearity of the subject-specific models. A Bayesian nonlinear
mixed-effects (BNLME) model allows us to incorporate prior
information at the population level into the estimates of dynamic
parameters for individual subjects. We briefly summarize the main
concepts in the Bayesian approach to inference and the presentation
is, of course, far from exhaustive
[Davidian and Giltinan (\citeyear{Davidian95bib8}); Gelfand and Smith (\citeyear{Gelfand90bib11}); Huang et al.
(\citeyear{Huang06bib18}); Wakefield et al. (\citeyear{Wakefield94bib41}); Wakefield (\citeyear{Wakefield96bib42})].

In reference to the model (\ref{ode2}), we denote the number of
subjects by \textit{n} and the number of measurements on the \textit{i}th
subject by $ m_i$. Let ${\bolds{\mu}}=(\log c,\log\delta, \log d_T,
\log
\rho, \log R, \beta_0,\beta_1, \beta_2)^T$,
${\bolds{\Theta}}=\{{\bolds{\theta}}_i,i=1,\ldots,n\}$, $\bolds{\theta
}_\mathit{ i}=(\log c_i,\break\log\delta_i,\log d_{Ti}, \log\rho_i, \log R_{i}, \log\phi_i)^T$ and
$\mathbf{Y}=\{y_{ij},i=1, \ldots,n; j=1,\ldots,m_i\}$. Let
$f_{ij}(\bolds{\theta}_i, t_j)=\log_{10}(V(\bolds{\theta}_i,t_j))$, where
$V(\bolds{\theta}_i,t_j)$ is proportional to the numerical solution of
$\widetilde{V}(t)$ in the differential equations (\ref{ode2}) for
the \textit{i}th subject at time $\mathit{ t_j}$. Let $y_{ij}(t)$ and
$e_i(t_j)$ denote the repeated measurements of
viral load in $\log_{10}$ scale and a measurement error with mean zero,
respectively. Note that log-transformation of dynamic parameters and
viral load is used to make sure that estimates of dynamic parameters
are positive and to stabilize the variance and convergence,
respectively. The BNLME model can be written in the following three
levels [Gelfand and Smith (\citeyear{Gelfand90bib11}); Davidian and Giltinan (\citeyear{Davidian95bib8});
Huang and Wu (\citeyear{Huang06bib17}); Wakefield (\citeyear{Wakefield96bib42})].

\textit{Level} 1. Within-subject variation:
\begin{equation}
\mathbf{y}_i=\mathbf{f}_i({\bolds{\theta}}_i)+\mathbf{e}_i,\qquad
\mathbf{e}_i|\sigma^2, {\bolds{\theta}}_i \sim\mathcal{N}(\mathbf
{0},{\sigma}^{\mathrm{2}} \mathbf{I}_\mathit{{m_i}}),
\label{s1}
\end{equation}
where $\mathbf{y}_i=(y_{i1}(t_1),\ldots,y_{i m_i}(t_{m_i}))^T$,
$\mathbf{f}_{i}({\bolds{\theta}}_i)=(f_{i1}({\bolds{\theta
}}_i,t_1),\ldots,f_{i
m_i}({\bolds{\theta}}_i, t_{m_i}))^T$,
$\mathbf{e}_i=(e_{i}(t_1),\ldots,e_{i}(t_{m_i}))^T$.

\textit{Level} 2. Between-subject variation:
\begin{equation}
{\bolds{\theta}}_\mathit{ i}={\mathbf{W}}_i {\bolds{\mu}}+\mathbf{b}_i,\qquad
[\mathbf{b}_i|{\bolds{\Sigma}}] \sim\mathcal{N}({\mathbf{0}},{\bolds{\Sigma}}),
\label{s2}
\end{equation}
where $\mathbf{b}_i$ are random effects with mean zero. It is
noteworthy that, for $\bolds{\beta}=(\beta_0, \beta_1, \beta
_2)^T$, no
log-transformation is required as they are not necessarily positive.
${\mathbf{W}}_i=(\mathbf{I}_6, \mathbf{J}_{1i}, \mathbf{J}_{2i})$,
where $\mathbf{I}_6$ is
an identity matrix and $\mathbf{J}_{si}=(0,0,0,0,0, w_{si})^T$ $(s=1,2;
i=1,2,\ldots,n)$ are $6\times1$ vector, with ${w_{1i}}$ and
${w_{2i}}$ being (standardized) individual baseline viral load (in
$\log_{10}$ scale) and CD4 cell count, respectively. For
$\bolds{\beta}=(\beta_0, \beta_1, \beta_2)^T$, we are only
interested in
estimating them at population level. Thus, the individual parameter
$\phi_i$ is related to them as follows,
$\log\phi_i=\beta_0+\beta_1{w_{1i}}+\beta_2{w_{2i}}+b_{i6}$, where
$b_{i6}$ is the last element of $\mathbf{b}_i$ $(i=1,2,\ldots,n)$.

\textit{Level} 3. Hyperprior distributions:
\begin{equation}
\sigma^{-2} \sim\mathit{Ga}(a,b),\qquad
{\bolds{\mu}} \sim\mathcal{N}({\bolds{\eta}},\bolds{\Lambda}),\qquad
{\bolds{\Sigma}}^{\mathrm{-1}} \sim\mathit{Wi}({\bolds{\Omega}},\nu),
\label{s3}
\end{equation}
where the mutually independent Gamma ($\mathit{ Ga}$), Normal
($\mathcal{N}$) and Wishart ($\mathit{ Wi}$) prior distributions are
chosen to facilitate computations [Davidian and Giltinan (\citeyear{Davidian95bib8})]. The
hyper-parameters $a, b, {\bolds{\eta}},{\bolds{\Lambda}}, {\bolds \Omega}$
and $\nu$
can be determined from previous studies and literature.

The Bayesian approach is developed in the presence of observations
whose value is initially uncertain and described through a probability
distribution, which depends on some parameters. In the applications
we assume that the researcher has some knowledge about at least some
of the parameters which often represent characteristics of interest
describing the process. The Bayesian approach incorporates this
information through prior distribution into observed data to obtain
its posterior distribution. While computation of the posterior
distribution involves solving multidimensional integrals, the
introduction of Markov chain Monte Carlo (MCMC) methods such as the
Gibbs sampler and Metropolis--Hastings algorithm opened the way to
analysis of complex models through decomposition and sampling from
full conditional distributions; see Gamerman (\citeyear{Gamerman97bib10}) and Gilks et al.
(\citeyear{Gilks95bib11-0}) for general theory and implementation details. Some more
specific discussion of the Bayesian dynamic modeling approach,
including the choice of the hyper-parameters, the iterative MCMC
algorithm and the implementation of the MCMC procedures can be found
in publications by Huang and Wu (\citeyear{Huang06bib17}) and Wakefield (\citeyear{Wakefield96bib42}). The
Bayesian approach was developed and tailored as required by the
unique features of the proposed HIV dynamic models. The basic
principles of these proposed methodologies were well established in
the statistical literature [Gamerman (\citeyear{Gamerman97bib10}); Gilks et al. (\citeyear{Gilks95bib11-0});
Wakefield (\citeyear{Wakefield96bib42})], but the applications of these methods in this
paper are nonetheless innovative within the context of a system of
nonlinear ODE of time-varying coefficient, but without a closed-form
solution.

The progress in Bayesian posterior
computation due to MCMC procedures has made it possible to fit
increasingly complex statistical models [Huang and Wu (\citeyear{Huang06bib17});
Wakefield (\citeyear{Wakefield96bib42})] and entailed the wish to determine the best fitting
model in a class of candidates. Thus, it has become more and more
important to develop efficient model selection criteria. A recent
publication by Spiegelhalter {et al.} (\citeyear{Spiegelhalter02bib32}) suggested a
generalization of the Akaike information criterion (AIC) [Akaike
(\citeyear{Akaike73bib2})] and related also to the Bayesian information criterion (BIC)
[Schwarz (\citeyear{Schwarz78bib35})] that is deviance information criterion (DIC).
In this paper we demonstrate its usefulness to compare BNLME
models for longitudinal HIV dynamics discussed previously. For
completeness, a brief summary of DIC follows. More detailed
discussion of DIC and its properties
can be found in publications by Spiegelhalter et al. (\citeyear{Spiegelhalter02bib32}) and Zhu and
Carlin (\citeyear{Zhu00bib48}).

Assume that the distribution of the data, $\mathbf{Y}$, depends
on the parameter vector~$\bolds{\Psi}$. Most recently, Spiegelhalter
{et al.}
(\citeyear{Spiegelhalter02bib32}) suggested examining the posterior distribution of the
deviance statistics defined by
\[
D(\bolds{\Psi})=-2\log p(\mathbf{Y}|\bolds{\Psi})+2\log g(\mathbf{Y})
\]
for Bayesian model comparison, where $p(\mathbf{Y}|\bolds{\Psi})$
is the
likelihood function, that is, the conditional joint
probability density function of the observed data $\mathbf{Y}$ given
the parameter vector $\bolds{\Psi}$, and $g(\mathbf{Y})$ denotes a fully
specified standardizing term that is a function of the data alone
(which thus has no impact on model selection). Based on the
posterior distribution of $D(\bolds{\Psi})$, DIC consists of two components
as follows:
\begin{equation}
\mathit{DIC}=\bar{D}+p_D=2\bar{D}-D(\bar{\bolds{\Psi}}),
\label{DIC}
\end{equation}
where $\bar{D}=E_{\bolds{\Psi}|\mathbf{Y}}[D(\bolds{\Psi
})]=E_{\bolds{\Psi}|\mathbf{Y}}[-2\log
p(\mathbf{Y}|\bolds{\Psi})]$ and $p_D=\bar{D}-D(\bar{\bolds{\Psi
}})$ is the effective
number of parameters, defined as the difference between the
posterior mean of the deviance and the deviance evaluated at the
posterior mean $\bar{\bolds{\Psi}}$ of the parameters. As with other model
selection criteria, we caution that DIC is not intended for
identification of the `correct' model, but rather merely as a method
of comparing a collection of alternative formulations. In our model
with different baseline characteristics and/or the MEMS adherence
summary metrics, DIC can be used to identify the most significant
covariate and MEMS adherence summary metrics in contribution to
virologic response. Under the model (\ref{s1}), in the absence of
any standardizing function $g(\mathbf{Y})$, the deviance is
\begin{eqnarray}\label{dev}
D(\sigma^{-2},{\bolds \Theta})&=&-2 \log p(\mathbf{Y}|\sigma^{-2},\bolds
{\Theta})
\nonumber
\\[-8pt]
\\[-8pt]
\nonumber
&=& \sum_{i=1}^n \sigma^{-2}\bigl(\mathbf{y}_i-\mathbf{f}_i({\bolds
{\theta}}_i)\bigr)^T
\bigl(\mathbf{y}_i-\mathbf{f}_i({\bolds{\theta}}_i)\bigr)
- \log\sigma^{-2} \sum_{i=1}^n m_i.
\end{eqnarray}
As discussed above, our MCMC approach to estimating DIC first draws
$\{(\sigma^{-2}, {\bolds \Theta})^{(g)}\}_{g=1}^G$ values from the
posterior distribution, and then calculates corresponding
$\{D^{(g)}\}_{g=1}^G$ values from (\ref{dev}), where $G$ is the
number of samples of posterior distribution. Finally, we estimate
DIC as $2\bar{D}- D(\bar{\sigma}^{-2},{\bar{\bolds{\Theta}}})$, where
$\bar{D}=\frac{1}{G}\sum_{g=1}^{G}D^{(g)}$,
$\bar{\sigma}^{-2}=\frac{1}{G}\sum_{g=1}^{G}(\sigma^{-2})^{(g)}$,
$\bar{\theta_i}=\frac{1}{G}\sum_{g=1}^{G}\theta_i^{(g)}$ and
$\bar{\bolds \Theta}=\{\bar{\theta_i}, i=1,\ldots,n\}$.

\section{Analysis of AIDS clinical data}\label{sec3}
\subsection{Motivating application and observed data} \label{sec3.1}

The subject sample in our analysis was drawn from the AIDS Clinical
Trials Group (ACTG) 398 study, a~randomized, double-blind, placebo-controlled, 4-Arm trial study of
amprenavir (APV) as part of several dual protease inhibitor (PI)
regimens in subjects with HIV infection in whom initial PI therapy
had failed. Subjects in all arms received APV (PI), three reverse
transcriptase inhibitors (RTI): efavirenz (EFV), abacavir (ABC) and
adefovir dipivoxil (ADV) plus a second PI or placebo: Arm~A
(saquinavir${}={}$SQV), Arm B (indinavir${}={}$IDV), Arm C (nelfinavir${}={}$NFV) and
Arm~D (placebo matched for one of these three PIs). Subjects are
HIV-infected individuals with prior exposure to approved PIs and who
have exhibited loss of virologic suppression as reflected by a
plasma HIV-1 RNA concentration of $\geq$1000 copies/ml. Subjects
were scheduled for follow-up visits at study (day 0), at weeks~2, 4,
8, 12, 16 and every 8 weeks thereafter until week 72, and at the
time of confirmed virologic failure. More detailed descriptions of
this study and data are given by Hammer et
al. (\citeyear{Hammer02bib13}) and Pfister et al. (\citeyear{Pfister03bib30}).

As indicated previously, the primary objective of this paper is to
investigate the effect of
adherence interaction with drug resistance and baseline covariates
to prescribed ARV therapy on virologic response measured repeatedly
over time in HIV-infected patients. We construct a novel HIV dynamic
model (which is parameter identifiable) with consideration of drug
adherence assessed by use of MEMS data, drug susceptibility
($\mathit{IC}_{50}$) and baseline covariates to link plasma drug
concentration to the long-term changes in HIV-1 RNA observation
after initiation of therapy. In the model we incorporate the two
clinical factors (drug adherence measured by MEMS data and drug
susceptibility) and baseline viral load and CD4 cell count into a
function of treatment efficacy (see Section \ref{sec2.2}).

Because phenotype sensitivity testing was performed only on a subset
of randomly selected subjects, the number of subjects available for
our analysis was greatly reduced. We chose to consider only the
subjects within Arm C for our analysis because this arm afforded the
greatest number of subjects ($n=31$) with available phenotypic drug
susceptibility data on the two PIs (APV and NFV) and had available
MEMS adherence data, as required for our model. A summary of
measurements of data to be used in our analysis is briefly described
below.
%
\begin{figure}[b]
\vspace{-7pt}
\includegraphics{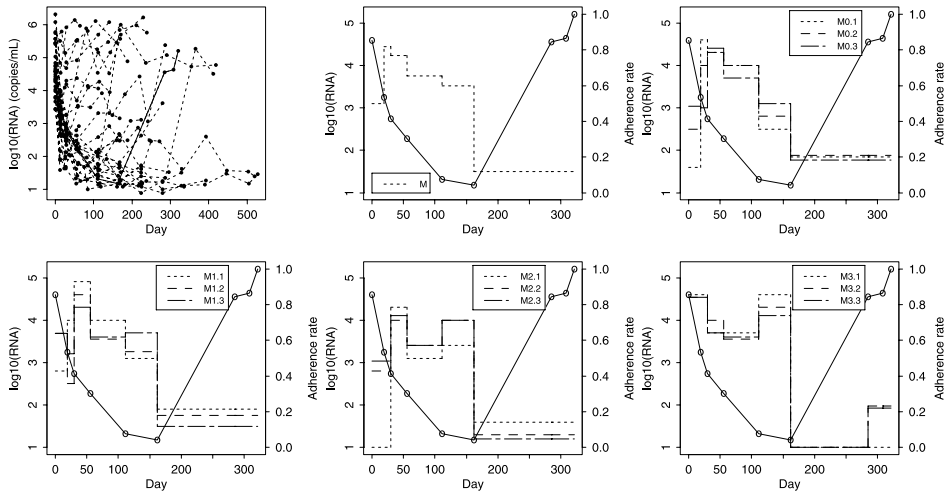}

\caption{The profiles of viral load measurements (in $\log_{10}$
scale) from the 31 patients (up-left panel) and one trajectory of
viral load (solid curve) and associated adherence rates (stairsteps)
over time from the thirteen summary measures of MEMS data with the
APV drug for one subject from ACTG398.} \label{fig1}
\vspace{-7pt}
\end{figure}

\textit{Plasma viral load}: Plasma viral load was measured in
copies/ml at designed study time by the ultrasensitive
reverse transcriptase--polymerase chain reaction HIV-1 RNA assay (Roche
Molecular Systems).
Only measurements taken while on protocol-defined treatment were used
in the analysis.
The exact day of viral load measurement (not predefined study week)
was used to compute study day in our
analysis. A $\log_{10}$ transformation was used in the analysis of
viral load data. The graph in Figure~\ref{fig1} (up-left panel) shows
the viral load trajectories of those subjects. Note that some viral
load measurements at designed study time were not observed due to
laboratory and other problems (for example, viral load measurement
was not observed at week 12 for the subject displayed in
Figure~\ref{fig1}).

\textit{Medication adherence}: Medication adherence was measured
by two methods:
by the use of questionnaires and by the use of MEMS [Pfister et al. (\citeyear{Pfister03bib30})].
Subjects completed an adherence questionnaire at study weeks. The questionnaire
was completed by the study participant and/or by a face-to-face
interview with study personnel. For MEMS, an MEMS cap was used to
monitor APV and EFV compliance only.
The subjects were asked to bring their
medication bottles and caps to the clinic at each study visit, where
cap data were downloaded to computer files and stored for later
analysis. The MEMS adherence rate for APV was determined as the sum
of positive dosing events divided by the sum of prescribed dosing
events during the specified time interval. In our analysis, we
assumed that NFV had the same MEMS adherence rate as APV since both
APV and NFV were prescribed with the same dosing schedule (twice
daily), a prescribed AM and PM dosing period was defined for each
subject and, hence, the bottles were opened twice per day [Pfister et
al. (\citeyear{Pfister03bib30})]. As discussed previously, this study focuses mainly
on investigating optimal strategy to summarize adherence rates
determined by MEMS data for efficient production of virologic
responses. For the MEMS data analysis, it was not possible to model
daily adherence rates and instead the adherence rate was computed
with the following scenarios to consider effects of both interval
length and time frame (delay of timing) for MEMS assessment.

%
\begin{table}[b]
\vspace{-7pt}
\tabcolsep=0pt
\caption{Summary of the MEMS interval definitions and other
information} \label{tab1}
\begin{tabular*}{\textwidth}{@{\extracolsep{\fill}}lcccc@{}}
\hline
& & \multicolumn{2}{c}{\textbf{Adherence interval definition}}& \\[-6pt]
& & \multicolumn{2}{c}{\hrulefill}& \\
\textbf{Case} &\textbf{MEMS adherence} & \textbf{Time frame length} & \textbf{Interval} &
\textbf{Example for week 8}\\
&\textbf{interval name} & \textbf{(weeks prior to viral load} & \textbf{length} & \textbf{(day 56), adherence} \\
& &\textbf{measurement)}&& \textbf{computed over days}\\
\hline
\phantom{0}1& M & 0 week & visit time& 28--55 \\
\phantom{0}2& M0.1 & 0 week & 1 week & 49--55 \\
\phantom{0}3& M0.2 & 0 week & 2 weeks & 42--55 \\
\phantom{0}4& M0.3 & 0 week & 3 weeks & 35--55 \\
\phantom{0}5& M1.1 & 1 week & 1 week & 43--49 \\
\phantom{0}6& M1.2 & 1 week & 2 weeks & 36--49 \\
\phantom{0}7& M1.3 & 1 week & 3 weeks & 29--49 \\
\phantom{0}8& M2.1 & 2 weeks & 1 week & 36--42 \\
\phantom{0}9& M2.2 & 2 weeks & 2 weeks & 29--42 \\
10& M2.3 & 2 weeks & 3 weeks & 22--42 \\
11& M3.1 & 3 weeks & 1 week & 29--35 \\
12& M3.2 & 3 weeks & 2 weeks & 22--35 \\
13& M3.3 & 3 weeks & 3 weeks & 15--35 \\
\hline
\end{tabular*}
\vspace{-7pt}
\end{table}

To determine the best summary metric of the MEMS adherence rate, we
evaluated different assessment interval lengths (averaging adherence
dosing events over~1, 2 or 3 week intervals) and different
assessment time frames (fixing the assessment interval times to end
either immediately or 1, 2 or 3 weeks prior to the next measured
viral load). Table~\ref{tab1} summarizes the MEMS assessment
interval notation and definitions for the 13 scenarios. As an
example, M2.2 in Table~\ref{tab1} denotes an MEMS adherence interval
length of 2 weeks fixed to end 2 weeks prior to the next viral load
measurement; for instance, the MEMS adherence rate for a subject at
study week 8 (day 56) was calculated as the number of nominal dosing
events divided by the number of prescribed dosing events over study
days 29--42 and this value was used to represent adherence from the
previous study visit to the study visit at the day 56 for modeling.
The case M serves as a reference and averages all the available MEMS
data between viral load measurements. As an example, the viral load
(in $\log_{10}$ scale) and adherence rates over time from the
thirteen cases of MEMS data with APV drug for the one representative
subject are presented in Figure~\ref{fig1}.

%
%
\begin{figure}
\vspace{-7pt}
\includegraphics{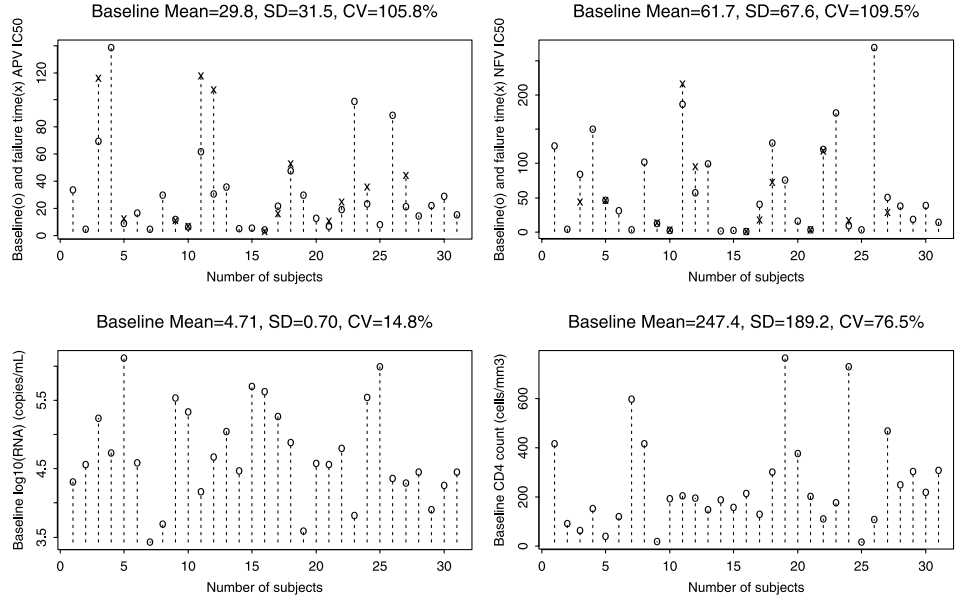}

\caption{The baseline ($\circ$) and failure time ($\times$) $\mathit{IC}_{50}$
for APV/NFV drugs (upper panel) with baseline $\mathit{IC}_{50}$ mean,
standard deviation (SD) and coefficient of variation (CV),
respectively, and the baseline viral load in $\log_{10}$ scale and
baseline CD4 cell count (lower panel) with mean, SD and CV,
respectively, for the 31 subject-specific individuals from the ACTG398
study. Note that for a subject, if a single measurement of $\mathit{IC}_{50}$
is observed at baseline only, there is no $\times$ sign appearing in
the upper panel of the plot.}\vspace*{-4pt}
\label{fig2}
\vspace{-7pt}
\end{figure}

\textit{Phenotypic virus susceptibility to drug}: The phenotypic
virus resistance to drug were retrospectively determined from
baseline samples. Patients were selected to have samples assayed
based on receiving study treatment for at least 8 weeks and having
available sample. Some patients had virologic failure and
phenotypic susceptibility
testing done on samples at the time of failure. Testing was done via
the recombinant virus assay
(PhenoSense, ViroLogic Inc., South San Francisco, CA). For analysis, we
used the phenotype
marker, $\mathit{IC}_{50}$ [Molla et al. (\citeyear{Molla96bib23})], to quantify agent-specific drug
resistance. We refer to this marker as the median inhibitory
concentration. The baseline ($\circ$) and failure time ($\times$)
$\mathit{IC}_{50}$'s of 31 subject-specific individuals for the APV and NFV
drugs are displayed in Figure~\ref{fig2} (upper panel) and are used to
construct $\mathit{IC}_{50}(t)$. Note that some subjects have only baseline
$\mathit{IC}_{50}$ due to the fact that they maintained viral
suppression or dropped out from the study. If no $\mathit{IC}_{50}$
measurement is observed at failure time for a subject, $\mathit{IC}_{50}(t)$
becomes a constant in this case.

\textit{Baseline characteristics}: The baseline
viral load in $\log_{10}$ scale (VL) and the baseline CD4 cell count
were chosen as covariates in the model for data analysis.
The log-transformation of viral load is used to
stabilize the variance of measurement error and estimation
algorithm. The baseline characteristics of 31 subject-specific
individuals with mean, standard deviation (SD) and coefficient of
\mbox{variation} (CV) are displayed in Figure~\ref{fig2} (lower panel). To
avoid very small (large) estimates which may be unstable, we
standardized these covariate values. For baseline $\log_{10}$(RNA),
for instance, each $\log_{10}$(RNA) value is subtracted by mean
(4.71)
and divided by standard deviation (0.70).

\subsection{Model fitting and parameter estimation results}
\label{sec3.2}
In this section we apply the BNLME modeling approach
to fit the data described in Section \ref{sec3.1}. Based on the
discussion in Section \ref{sec2}, the prior distribution for
${\bolds{\mu}}$ was assumed to be $ \mathcal{N}({\bolds{\eta
}},{\bolds \Lambda})$ with
${\bolds \Lambda}$ being a diagonal matrix. Following the idea of
Huang and Wu (\citeyear{Huang06bib17}) for prior construction, as an example we discuss
the prior construction for $\log\delta$. The prior constructions
for other parameters are similar and so are omitted here.

Ho et al. (\citeyear{Ho95bib15}) reported viral dynamic data on 20 patients; the
logarithm of the average death rate of infected cells ($\log
\delta$) is $-$1.125. Wei et al. (\citeyear{Wei95bib42-1}) used two different models
with a group of 22 subjects to estimate death rate of infected cells
and obtained $\log\delta$ with $-0.84$ and $-1.33$, respectively.
Following these two studies, Nowak et al. (\citeyear{Nowak95bib24}) estimated $\log
\delta= -0.934$ based on 11 subjects with one possible outlying
subject excluded. It can be seen that four estimates of $\log
\delta$ from these studies are $-1.125$, $-0.84$, $-1.33$ and $-0.934$,
respectively. The individual estimates of $\log\delta$ from these
studies approximately follow a symmetric normal distribution. Thus,
we chose a normal distribution $\mathcal{N}$($-1.0$, 100.0) as the
prior for $\log\delta$ (the large variance indicated that we used a
noninformative prior for $\log\delta$). Similarly, the values of
the hyper-parameters at population level are chosen as follows [Ho
et al. (\citeyear{Ho95bib15}); Nowak et al. (\citeyear{Nowak95bib24}); Nowak and May (\citeyear{Nowak00bib25}); Perelson et
al. (\citeyear{Perelson96bib27,Perelson97bib28});
Perelson and Nelson (\citeyear{Perelson99bib29}); Verotta (\citeyear{Verotta05bib36}); Wei et al. (\citeyear{Wei95bib42-1})]:
\begin{eqnarray*}
a&=&4.5,\qquad b=9.0, \qquad\nu=10.0,\\
\bolds{\Lambda}&=&{\operatorname{diag}}(100.0, 100.0, 100.0, 100.0, 100.0, 100.0, 100.0,
100.0),\\
\bolds{\eta}&=&(1.1, -1.0, -2.5, 1.2, 1.0, 1.0, 0.5, 0.5)^T,\\
\bolds{\Omega}&=&{\operatorname{diag}}( 2.5, 2.5, 2.5, 2.5, 2.5, 2.5).
\end{eqnarray*}

We decide that one long chain is run for MCMC implication with
considerations of the following two issues: (i) a number of initial
``burn-in'' simulations are discarded, since from an arbitrary
starting point it would be unlikely that the initial simulations
came from the stationary distribution targeted by the Markov chain;
(ii) one may only save every $k$th ($k$ being an integer) simulation
sample to reduce the dependence among samples used for parameter
estimation. Because the antiviral response modeling involves
numerically solving nonlinear differential equations, thus
computational burdens would be more pronounced with the Bayesian
approaches via MCMC procedure. Utilizing efficient computer
algorithms are critical in this regard. Therefore, we are going to
adopt these strategies in our MCMC implementation using FORTRAN code
that calls a differential equation subroutine solver (DIVPRK) in the
IMSL library (\citeyear{IMSL94bib20}), which uses the Runge--Kutta--Verner fifth-order
method. The computer codes are available from the corresponding
author upon request. An informal check of convergence is conducted
based on graphical techniques according to the suggestion of Gelfand
and Smith (\citeyear{Gelfand90bib11}). Based on the results, we propose that, after an
initial number of 20,000 burn-in iterations, every 4th MCMC sample
was retained from the next 80,000 samples. Thus, we obtained 20,000
samples of targeted posterior distributions of the unknown parameters.

We fitted the model to the data from 31 subjects discussed in
Section \ref{sec3.1} using the proposed BNLME modeling approach. We
incorporate the two clinical factors, drug adherence assessed by
MEMS cap data and drug susceptibility (phenotype $\mathit{IC}_{50}$ values),
as well as baseline covariates into a function of drug efficacy. For
model fitting, adherence rates were determined from MEMS data with
13 different scenarios. For model fitting and the purpose of
comparisons, we set up a control model as the one without using any
drug adherence, resistance and covariate information which
corresponds to setting $\gamma(t)=2/(\exp(\beta_0)+2)$ with
$\mathit{IC}_{50}(t)=1$, $A(t)=1$ and $w_1=w_2=0$; for this case, our model
reverts to that discussed by Nowak and May (\citeyear{Nowak00bib25}) and Perelson and
Nelson (\citeyear{Perelson99bib29}). The other 13 models are specified based on the
combination of drug resistance ($\mathit{IC}_{50}$), baseline covariate data
and 13 different adherence summary metrics listed in
Table~\ref{tab1}.

%
\begin{figure}
\vspace{-7pt}
\includegraphics{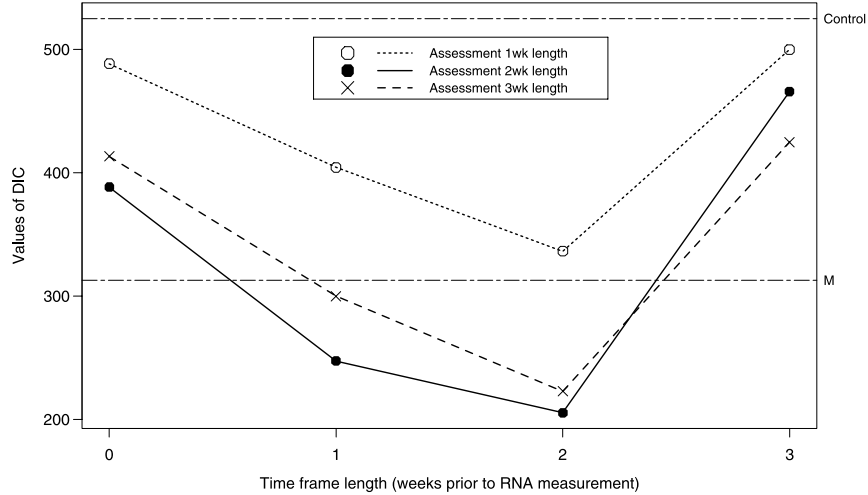}

\caption{Comparison of the DIC values for the models from 13
different determinants of MEMS adherence, interacted by drug
resistance and covariates, with the control model. The two
horizontal lines represent the DIC values for the control model and
the reference model with adherence rate determined by case M,
respectively.} \label{dic}
\vspace{-7pt}
\end{figure}

%
\begin{figure}[b]
\vspace{-7pt}
\includegraphics{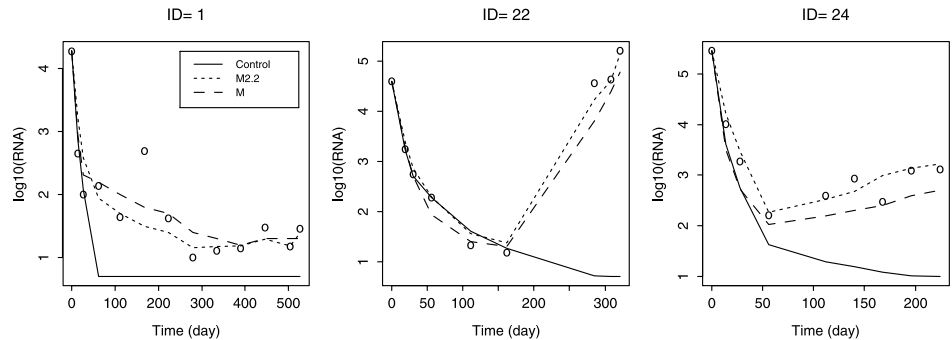}
\caption[Illustration ]{The estimate of viral load trajectory from
the model fitting with the 3 different determinants of adherence:
\textup{(i)} Control model (solid curves), \textup{(ii)} M2.2 model (dotted curves)
and \textup{(iii)} reference~(M) model (dashed curves) for the three
representative subjects. The observed values are indicated by
circles.}
\vspace{-7pt}
\label{fit}
\end{figure}

In order to assess how adherence rates, determined from 13 different
scenarios, interacted with drug susceptibility and covariates to
contribute to~virologic response, we fitted the models to all 13
scenarios as well as the control model and compared the fitting
results. We found based on the DIC criterion (see Figure~\ref{dic})
that, overall, the model with adherence rate determined from
MEMS dosing events, taken time frame length of 2 weeks prior to a
viral load measurement with over either a 2 week assessment interval
(M2.2) or~a 3 week assessment interval (M2.3), provided the best fits
to the observed data, compared to the other 12 models for most
subjects. The reference model with adherence rate averaged by all
the available MEMS data between viral load measurements gave a
moderate fit to the observed data. We clearly see that all models
fit the early viral load data well, but~the control model, lacking
factors for subject-specific drug adherence and susceptibility as
well as baseline covariates, failed to fit viral load rebounds and
fluctuations, and provided the worst fitting results for the
majority of subjects. For the purpose of illustration, the model
fitting curves from the control model (solid curves), the best fit
model (M2.2: dotted curves) and the reference model (M: dashed
curves) are displayed in Figure~\ref{fit} for the three representative
subjects.

\begin{figure}
\vspace{-7pt}
\includegraphics{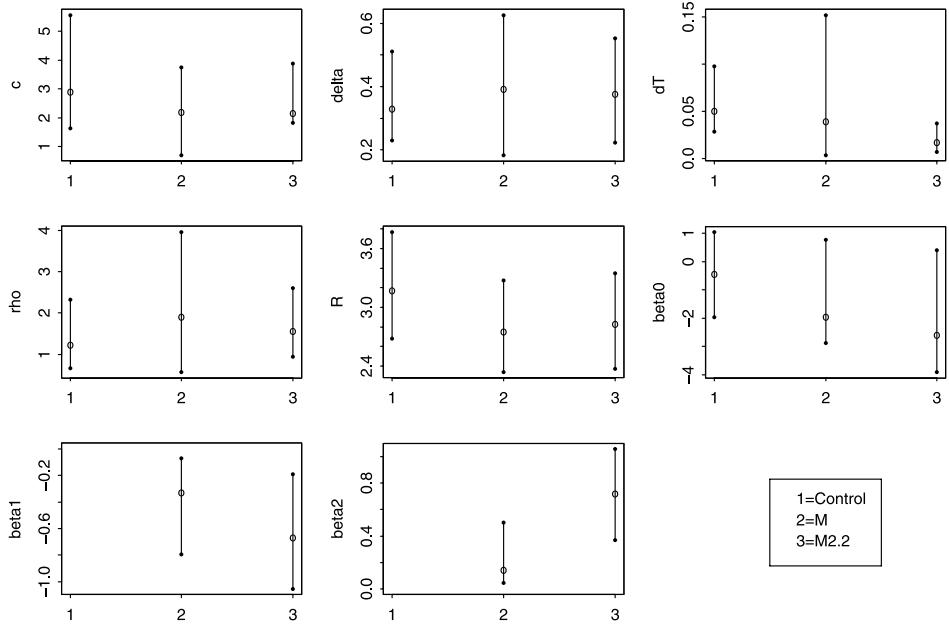}

\caption{A summary of the estimated posterior means ($\circ$) of
population parameters and the corresponding $95\%$ equal-tail
credible intervals (CI) for the models from 3 different determinants
of adherence.} \label{par}
\vspace{-7pt}
\end{figure}

For the purpose of comparison, Figure~\ref{par} presented the
population \mbox{posterior} means and the corresponding 95\% equal-tail
credible intervals (CI) of the~eight parameters for the control
model, the best fit model (M2.2) and the reference model (M). For the
six dynamic parameters $(c, \delta, d_T, \rho, R, \beta_0)$, it is~shown
that the population estimates for the control model have
higher clearance rate of free virions~($c$), lower death rate of
infected cells ($\delta$), higher death~rate of target $T$ cells~($d_T$),
smaller viral load scaling factor ($\rho$), higher basic
repro\-ductive ratio for the virus ($R$) and larger $\phi$ than those
for the best~fit model (M2.2) and reference model (M), while the
population estimates for the best fit model and reference model are
generally similar. These differences may result from the effects of
drug adherence interacted with drug resistance and covariates in the
models. For the other two covariate effect parameters $(\beta_1,
\beta_2)$ which are relevant to treatment effect, we will discuss
them sepa\-rately in Section~\ref{sec3.4}. In terms of the
individual parameter estimates, a large between-subject variation in
the estimates of all individual dynamic parameters was observed
(data not shown here). Overall, the coefficient of variation ranges
from 15.4\% to 88.9\% for all parameters.
%

\subsection{Effects of adherence rates determined by different
MEMS summary metrics} \label{sec3.3}

Figure~\ref{dic} in Section \ref{sec3.2} displayed a comparison of the
DIC values for the models from 13 different determinants of MEMS
adherence, interacted by drug resistance and covariates, with the
control model. The observed patterns shown in Figure~\ref{dic}
provided information to answer the following questions: (i) what
MEMS assessment interval length is best and (ii) what MEMS
assessment time frame (delay
effect of timing) is best?

We can see that when the time frame for
MEMS assessment is fixed, models with a 2 week MEMS assessment
interval length generally outperform models with an assessment
interval length of 1 or 3 weeks except for the time frame \mbox{length
with} 3 weeks prior to viral load measurement where the model with a
3~week MEMS assessment interval length performs best.

Regardless of the assessment interval length, models which assess
compliance 2~weeks
prior to viral load measurement generally outperform models which
assess compliance immediately before viral load measurement, 1 week
before or 3~weeks before viral load measurement. Overall, the model
with a MEMS assessment interval length of 2 weeks measured from 4 to
2~weeks prior to viral load measurement (M2.2) was significantly a
better predicator of viral load over time than any other models,
with the exception of the M2.3 model which shows no significant
difference from the M2.2 model in terms of DIC values.

\subsection{Treatment effects of baseline characteristics
interacted with clinical factors} \label{sec3.4}
\begin{figure}
\vspace{-7pt}
\includegraphics{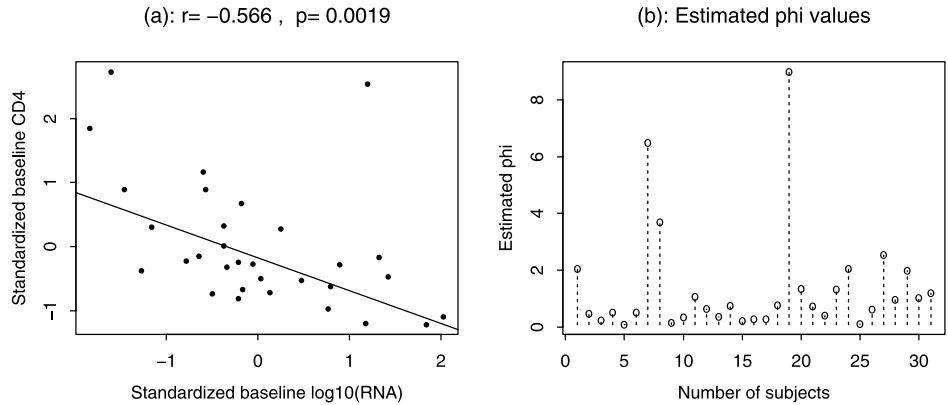}

\caption{\textup{(a)} Correlation between (standardized) baseline
$\log_{10}$(RNA) and CD4 cell count. The correlation coefficient
($r$) and p-value are obtained from the Spearman rank correlation test.
The line is a robust (MM-estimator) linear regression fit. \textup{(b)}
Estimated $\phi$ values for the 31 subject-specific individuals.}
\label{covariate}
\vspace{-7pt}
\end{figure}

Figure~\ref{par} summarized the population posterior means and the
corresponding 95\% equal-tail CI of the covariate effect parameters
$\beta_1$ and $\beta_2$ for the best fit model (M2.2) and the
reference model (M). It can be seen that estimates of $\beta_1$
(coefficient of baseline viral load) are negative, while estimates
of $\beta_2$ (coefficient of baseline CD4 cell count) are positive.
In fact, other models also provided the same scenarios for the
estimates of these two parameters (not shown here). As an example,
we report results based on the best fit model (M2.2). We can observe
from Figure~\ref{par} that the estimates of $\beta_1$ and $\beta_2$
are $\hat{\beta}_1=-0.67$ with $95\%$ CI ($-$1.056, $-$0.193) and
$\hat{\beta}_2=0.719$ with $95\%$ CI (0.371, 1.058). It indicates
that, according to antiviral drug efficacy model (\ref{eff}), the
baseline viral load ($\hat{\beta}_1=-0.67$) has a significant
positive effect on drug efficacy $\gamma(t)$, while the baseline CD4
cell count ($\hat{\beta}_2=0.719$) has a significant negative effect
on $\gamma(t)$ since the corresponding $95\%$ credible intervals do
not contain zero for both parameters. These findings could suggest
us with the following different ways. The lowest value of
$\gamma(t)$ [highest $\phi$ as displayed in Figure~\ref{covariate}(b)]
occurs in the subjects with the best prognosis (higher baseline CD4
cell count and lower baseline viral load). Alternatively, the
highest value of $\gamma(t)$ (lowest $\phi$) occurs in those with
the worst prognosis (lower baseline CD4 cell count and higher
baseline viral load). A possible explanation is that there is a
floor effect of viral load (or ceiling/floor effect of CD4 cell
count) that is not captured in the model. Further, given that
baseline CD4 cell count and viral load are jointly used to make
treatment decisions and are known to be negatively correlated as shown
in Figure~\ref{covariate}(a), the result based on the combination of
baseline viral load and CD4 count in $\gamma(t)$ indicates that the
baseline CD4 cell count and viral load have the opposite effect on
drug efficacy which might be intuitively understandable.

%
%

\section{A simulation study}\label{sec4}

In this paper we investigated the association between virologic
outcomes and medication adherence with confounding factors based on
the data from 31 subjects. As both one referee and the associate
editor suggested that a simulation study may be useful to evaluate
how our method performs, in this section we conduct a limited
simulation study here due to intensive computations involved. The
scenario we
consider is as follows.

We simulate a clinical trial with 31 patients receiving
antiviral treatment. For each patient, we assume
that the designs of this experiment, in particular, the sampling times
for viral load,
were the same as those in the ACTG 398 study.
The data for the phenotype marker (baseline and failure $\mathit{IC}_{50}$'s),
medication adherence and the baseline viral load/CD4 cell count were
taken from the ACTG 398 study, where medication adherence was
calculated by the M2.2 summary measure. The ``true'' values of
unknown parameters were the same as those estimated from the data set
of 31 subjects which were reported in Section \ref{sec3}. With
generated individual true parameters based on the equation
(\ref{s2}), known data [$\mathit{IC}_{50}(t), A(t), w_1$ and $w_2$], we
generated random samples for response (viral load) based on
model (\ref{ode2}). The values of hyper-parameters are chosen to be
the same as those in Section \ref{sec3}. For each simulated data set,
we fit the model using the Bayesian approach. The MCMC techniques
consisting of a series of Gibbs sampling and M--H algorithms were the
same as those in the real data analysis. We performed 50
replications and obtained the mean estimates (ME) of population
parameters together with the corresponding relative bias (RB), which
is the difference between the mean estimate and the true value of the
parameter divided by absolute value of the true parameter, and the
standard error (SE), defined as the square root of mean-squared
error divided by the absolute value of the true parameter.

%
%
\begin{table}
\vspace{-7pt}
\caption{The true values (TV) of parameters and mean estimates
(ME) of population parameters with 50 replications as well as the
corresponding relative bias (RB), defined as $100\times
(\mathit{ME}-\mathit{TV})/|\mathit{TV}|$, and standard error (SE), defined as
$100\times\sqrt{\mathit{MSE}}/|\mathit{TV}|$} \label{tab2}
\begin{tabular*}{\textwidth}{@{\extracolsep{\fill}}ld{2.3}d{2.3}d{2.3}d{2.3}@{}}
\hline
\multicolumn{1}{@{}l}{\textbf{Parameter}} & \multicolumn{1}{c}{\textbf{TV}} & \multicolumn{1}{c}{\textbf{ME}} &
\multicolumn{1}{c}{\textbf{RB ($\%$)}} & \multicolumn{1}{c@{}}{\textbf{SE ($\%$)}}\\
\hline
$\log c$ & 0.767 & 0.771 & 0.522 & 9.127 \\
$\log\delta$& -0.977 &-1.013 &-3.685 & 7.971 \\
$\log d_T$ & -4.086 &-4.101 &-0.367 & 5.871 \\
$\log\rho$ & 0.433 & 0.388 &-10.39 & 18.03 \\
$\log R$ & 1.040 & 1.100 & 5.769 & 3.089 \\
$\beta_0$ & -2.615 &-2.604 & 0.421 & 4.902 \\
$\beta_1$ & -0.670 &-0.665 & 0.746 & 10.13 \\
$\beta_2$ & 0.719 & 0.698 &-2.921 & 13.09 \\
\hline
\end{tabular*}
\vspace{-7pt}
\end{table}

In Table~\ref{tab2} we summarize the true values (TV) of parameters
and the ME of population parameters with 50 replications as well as
the corresponding RB and SE. The percentage is based on the absolute
value of the true parameter. It can be seen from Table~\ref{tab2}
that the RB ($\%$) for population parameter estimates are very
small, ranging from 0.522 to 10.39, and the SE ($\%$) ranges from
3.089 to 18.03. The simulation results indicate that our method with
considering the M2.2 model performs reasonably well in terms of
estimates of parameters except for the viral load proportionality
factor $\log\rho$ which has larger RB and SE. That is, our method
produces a substantially biased estimate and may severely
underestimate $\log\rho$. This may be explained by the fact that it
is probably caused from inaccurate numerical solutions to the system
of ODE (\ref{ode2}) which was used to construct the BNLME
model.\looseness=1

\section{Concluding discussion}\label{sec5}

In developing long-term dynamic modeling, this paper introduced a
dynamic mechanism specified by a system of time-varying ODE to (i)
establish a link between success of ARV therapy in virologic
response and MEMS adherence confounded by drug resistance and
baseline covariates, (ii)~fully integrate viral load, MEMS
adherence, drug resistance and baseline covariates data into the
statistical inference and analysis, and (iii) provide a powerful
tool to evaluate the effects of MEMS adherence determined by
a different summary metric on virologic response using the BNLME
modeling approach. This approach cannot only combine prior
information with current clinical data for estimating dynamic
parameters, but also deal with complex dynamic systems. Thus, the
results of estimated dynamic parameters based on this model should
be more reliable and reasonable to interpret long-term HIV dynamics.
Our models are simplified with the
main goals of retaining crucial features of HIV dynamics and,
at the same time, guaranteeing their applicability to typical
clinical data, in particular, long-term viral load measurements.
The proposed model fitted the clinical data reasonably well for
most patients in our study, although the fitting for a few patients
was not completely satisfactory because of unusual viral load
fluctuation patterns for these subjects.

We have explored the practical performance of DIC for the comparison
of developed models. DIC, a Bayesian version of the classical
deviance for model assessment, is particularly suited to compare
Bayesian models whose posterior distribution has been obtained
using MCMC procedures and can be used in complex hierarchical models
where the number of unknowns often exceeds the number of
observations and the number of free parameters is not well
defined. This is in contrast to AIC and BIC, where the number of free
parameters needs to be specified [Zhu and Carlin (\citeyear{Zhu00bib48})]. Overall,
combined with more traditional residual analysis and posterior
predictive model checks as discussed in this paper, DIC appears to
offer a comprehensive framework for comparison and evaluation within
a complex model class.

Several studies investigated the association between virologic
responses and adherence assessed by MEMS data only without
considering other confounding factors such as drug resistance using
standard modeling methods including Poisson regression [Knafl et
al. (\citeyear{Knafl04bib20-3})], logistic regression [Vrijens et al. (\citeyear{Vrijens05bib38})] and the linear
mixed-effects model [Liu et al. (\citeyear{Liu07bib21-2})]. In this article we employed
the proposed dynamic model and associated BNLME modeling approach to
assessment of effects of adherence determinants based on MEMS dosing
events in predicting virologic response. In particular, we
investigated (i) how to summarize the MEMS adherence data for efficient
prediction of virological response after accounting for potential
confounding factors such as drug resistance and baseline covariates,
and (ii) how to evaluate treatment effect of baseline
characteristics interacted with MEMS adherence and other clinical
factors. Note that a further study in comparing the performance of
these different methods may be important and warranted, although some
challenges are observed in terms of
different model structures and data characteristics.

The results indicate that the best summary metric for prediction of
virologic response based on DIC criterion is the adherence rate
determined by MEMS dosing events averaged over an assessment interval
of 2 or 3 weeks, and 2 weeks prior to the next measured viral load
observation (denoted by M2.2 or M2.3). We found that the best MEMS
adherence predictor (M2.2) of the effectiveness of ARV medications
on virologic response is consistent with that reported in Huang et
al. (\citeyear{Huang08bib18-2}) in which, however, the next best MEMS adherence predictor
(M1.2) is different from what is obtained in this paper. This
difference may be due to the various reasons as follows. In the
study by Huang et al. (\citeyear{Huang08bib18-2}), (i) it directly applied the model
(\ref{ode1}) to fit data and, thus, some assumptions were made due to
parameter unidentifiable issues; (ii) the analysis used the mean of
the sum of squared deviations as a criterion to evaluate model
fitting results; (iii) it assumed $\mathit{IC}_{50}$ data were extrapolated
linearly to the whole treatment period instead of a $\log$-linear
extrapolation offered in this paper which is considered more
reasonable biologically; and (iv) it did not incorporate baseline
covariates in the model. In addition, the superiority of the M2.2
model, associated with the MEMS adherence rate based on time frame length
of 2 weeks prior to a viral load measurement with over a 2 week
assessment interval, may be explained by the fact that it probably
reflects how long it takes for resistance mutations to first arise
and then come to dominate the plasma population of a virus. As pointed
out by an anonymous referee, this finding may also be interpreted as
follows. Low adherence two weeks prior to the viral load measurement
may not have had sufficient time for viral rebound to occur.

In this paper we set up a connection between subject-specific baseline
characteristics with interaction of clinical factors and drug
efficacy. We also found that, according to antiviral drug efficacy
model (\ref{eff}), the baseline viral load
had a positive effect on drug efficacy, while the baseline CD4 cell
count had a negative
effect on it. Our results may be explained by the fact that for
those patients with higher baseline viral load, the drug efficacy
needs to be higher than that for those with lower baseline viral
load. Therefore, a strong treatment is recommended for those
patients with higher baseline viral load. On the other hand,
patients with higher CD4 cell count may need lower drug efficacy so
that a more potent ARV drug regimen is not necessary for these
patients to avoid side-effects of drug use. The results may suggest
the benefit of initiating ARV therapy with a lower baseline viral
load and/or a higher baseline CD4 count. These results coincide
with those investigated by Notermans et al. (\citeyear{Notermans98bib23-1}) and Wu et al.
(\citeyear{Wu05bib46}) whose results were obtained using correlation analysis. Note
that given the estimated parameters, the subject with both a high
baseline CD4 cell count and a relatively high baseline viral load
[upper right quadrant of Figure~\ref{covariate}(a)] has a very
different $\phi$ than that with a similar baseline CD4 cell count,
but a low baseline viral load [upper left quadrant of
Figure~\ref{covariate}(a)]. It is possible that the subject in the
upper right quadrant was more recently infected (hence the higher
baseline CD4 cell count) or perhaps with a drug resistant virus and
would not be a candidate for a regimen with a ``less potent drug
efficacy.''

Our findings need to be interpreted in light of the study
limitations. First, in the ACTG 398 study, because phenotype
sensitivity testing was performed only on a subset of randomly
selected subjects, we chose 31 patients who have available data for
analysis in this paper. Second, due to reasons such as lost caps
and malfunction of caps, there were inaccurate MEMS data across the
treatment period which may not reflect actual adherence profile for
individual patients and, thus, the data quality could have some
impact on the results. Third, because of technical limitations,
the undetectable values of viral load were replaced with 25
copies/ml for analyses, which could introduce some bias due to a
cluster of ties of data points. Finally, this paper combined new
technologies in mathematical modeling and statistical inference with
advances in HIV/AIDS dynamics and ARV therapy to quantify complex
HIV disease mechanisms. The complex nature of HIV/AIDS ARV therapy
will naturally pose some challenges including missing data and
measurement error in clinical factors and covariates. These
complicated problems, which are beyond the scope of this article,
may be addressed, for example, using the joint model method [Wu
(\citeyear{Wu02bib47})] and other techniques [Carroll {et al.}  (\citeyear{Carroll95bib6})], and are warranted
for further investigation. Nevertheless, these limitations would not
offset the major findings from this study.

As the Associate Editor pointed out, we assumed that the
distributions of the random error and random effects
are normal, which is a common assumption in the literature for
statistical inference. However, due to the nature of AIDS clinical
data, it is possible that the data may contain outliers and/or
depart from normality and, thus, statistical inference and analysis
with normal assumption may lead to misleading results [Verbeke and
Lesaffre (\citeyear{Verbeke96bib35-1}); Ghosh et al. (\citeyear{Ghosh07bib11-1})]. Specially non-normal
characteristics such as skewness with heavy right or left tail may
appear often in virologic responses. Thus, a normality assumption
may be too restrictive to provide an accurate representation of the
structure that is often present in repeated measures and clustered
data. Thus, it is of practical interest to investigate nonlinear
models with a skew-normal distribution or $t$ distribution for
(within-subject) random error and random effects which are more
robust to outliers and skewness than those with a normal
distribution. In our recent study [Huang and Dagne (\citeyear{Huang10bib18-3})] we
addressed a Bayesian approach to nonlinear mixed-effects models in
conjunction with the HIV dynamic model and relaxed the normality
assumption by considering both random error and random-effects to
have a multivariate skew-normal distribution. The proposed model
provides flexibility in capturing a broad range of non-normal
behavior and includes normality as a special case. The results
suggest that it is very important to assume models with a
skew-normal distribution in order to achieve robust and reliable
results, in particular, when the data exhibit skewness. We are
actively applying this methodology into the data
investigated in this paper and will report the results in a future
study.

In summary, the mechanism-based dynamic model is powerful and
efficient to characterize relations between antiviral response and
medication adherence, drug susceptibility as well as baseline
characteristics, although some biological assumptions are required.
It is important to find a way to incorporate subject-specific
information with regard to drug susceptibility, medication adherence
and baseline characteristics in predicting long-term virologic
response. Since each of these factors may only contribute a very
small portion to virologic response and they may be confounded
through complicated interactions, the appropriate modeling of the
combination effects of these factors is critical to efficiently
utilize the information in virologic response predictions. The viral
dynamic model and associated statistical approaches discussed here
provide a good avenue to fulfill this goal. In particular, MEMS
adherence rate summarized by an optimal way in terms of assessing
both interval lengths and time frame lengths prior to viral load
measurement is an important factor that significantly determines the
effectiveness of ARV treatment and needs to be taken into account in
analysis of virologic responses. Our results demonstrate that MEMS
adherence data may not predict virologic response well unless the
MEMS cap data are summarized in an appropriate way as reported in
Section~\ref{sec3.3}. Additionally, although this paper
concentrated on HIV dynamics, the basic concept of longitudinal
dynamic systems and the proposed methodologies in this paper are
generally applicable to dynamic systems in other fields such as
biology, medicine, engineering or PK/PD studies as long as they
meet the relevant technical specification---a system of ODE.

\section*{Acknowledgments}

The authors are extremely grateful to the Editor, an Associate
Editor and one referee for their insightful comments and constructive
suggestions
that led to an improvement of the article. We gratefully acknowledge
ACTG 398 study
investigators for allowing us to use the clinical data from their study.
The authors are indebted to Dr. Susan L. Rosenkranz from Frontier
Science \& Technology Research Foundation
and SDAC of Harvard School of Public Health for her informative
discussions and data preparations.

%


\printaddresses

\end{document}